\documentclass{cjaa}                   

\usepackage{graphicx}                   
\input{epsf.sty}                        
\input{psfig.sty}                       

\begin{document}

\title{Applying Magnetized Accretion-Ejection Models to Microquasars:
a preliminary step}

\volnopage{Vol.0 (200x) No.0, 000--000}      
\setcounter{page}{1}          

\author{Petrucci P.O., Ferreira J., Tricot S., Pelletier G., Cabanac C.,
Henri G.}

\offprints{P.O. Petrucci}

\institute{Laboratoire d'Astrophysique de l'Observatoire de Grenoble, 414 rue
  de la Piscine, 38041 Grenoble Cedex 9, FRANCE\\
  petrucci@obs.ujf-grenoble.fr}


\date{Received~~2004~~~~~~~~~~~~~~~ ; accepted~~2004~~~~~~~~~~~~~~ }

\abstract{We present in this proceeding some aspects of a model that
should explain the spectral state changes observed in microquasars. In
this model, ejection is assumed to take place only in the innermost disc
region where a large scale magnetic field is anchored. Then, in opposite
to conventional ADAF models, the accretion energy can be efficiently
converted in ejection and not advected inside the horizon. We propose
that changes of the disc physical state (e.g. transition from optically
thick to optically thin states) can strongly modify the magnetic
accretion-ejection structure resulting in the spectral variability. After
a short description of our scenario, we give some details concerning the
dynamically self-consistent magnetized accretion-ejection model (Ferreira
1997; Casse \& Ferreira 2000) used in our computation. We also present
some preliminary results of spectral energy distribution.
\keywords{magnetohydrodynamics: MHD; radiation mechanisms: general;
X-rays: binaries}}

\authorrunning{Petrucci et al.}
\titlerunning{Applying Magnetized Accretion-Ejection Models to Microquasars}

\maketitle

%
%
\section{Introduction}           
\label{sect:intro}
Microquasars are X-ray binaries that present clear signatures of
transient or persistent jets. These objects exhibit also different
spectral states in X-ray. It is important to note that each state is
actually a stationary state in terms of dynamical time scales. Indeed,
they last about $10^6$ times longer. Thus any variation and sudden
transitions we observe can only be understood as slow (secular)
variations. The mechanism of these transitions has still to be found.

As a preliminary step of a more detailed work presently in progress, we
describe here the main characteristics of the model that we propose to
explain the spectral state changes observed in microquasars. This model
suppose the existence of a magnetized accretion-ejection structure
(hereafter MAES) that is developed in our group. These MAES well explain
the jet formation mechanism (e.g Ferreira 1997) but up to now the
emission processes were not included in the computation. We believe that
changes in the MAES (for example transition from optically thick to
optically thin accretion disc) could explain the spectral state changes
observed in galactic black holes and we present here the first sketch of
the radiative transfer treatment in such structures.

We present our scenario in Sect 2. In Sect. 3 we give some details of the
main characteristics of MAES used in our computations. In Sect 4, we show
preliminary disc SEDs in order to exemplify the effect of jet production
on the disc emission.

\section{Our Scenario}
\begin{figure}[b!]
    \begin{tabular}{cc}
      \hspace*{-4.5cm}\includegraphics[width=0.3\textwidth]{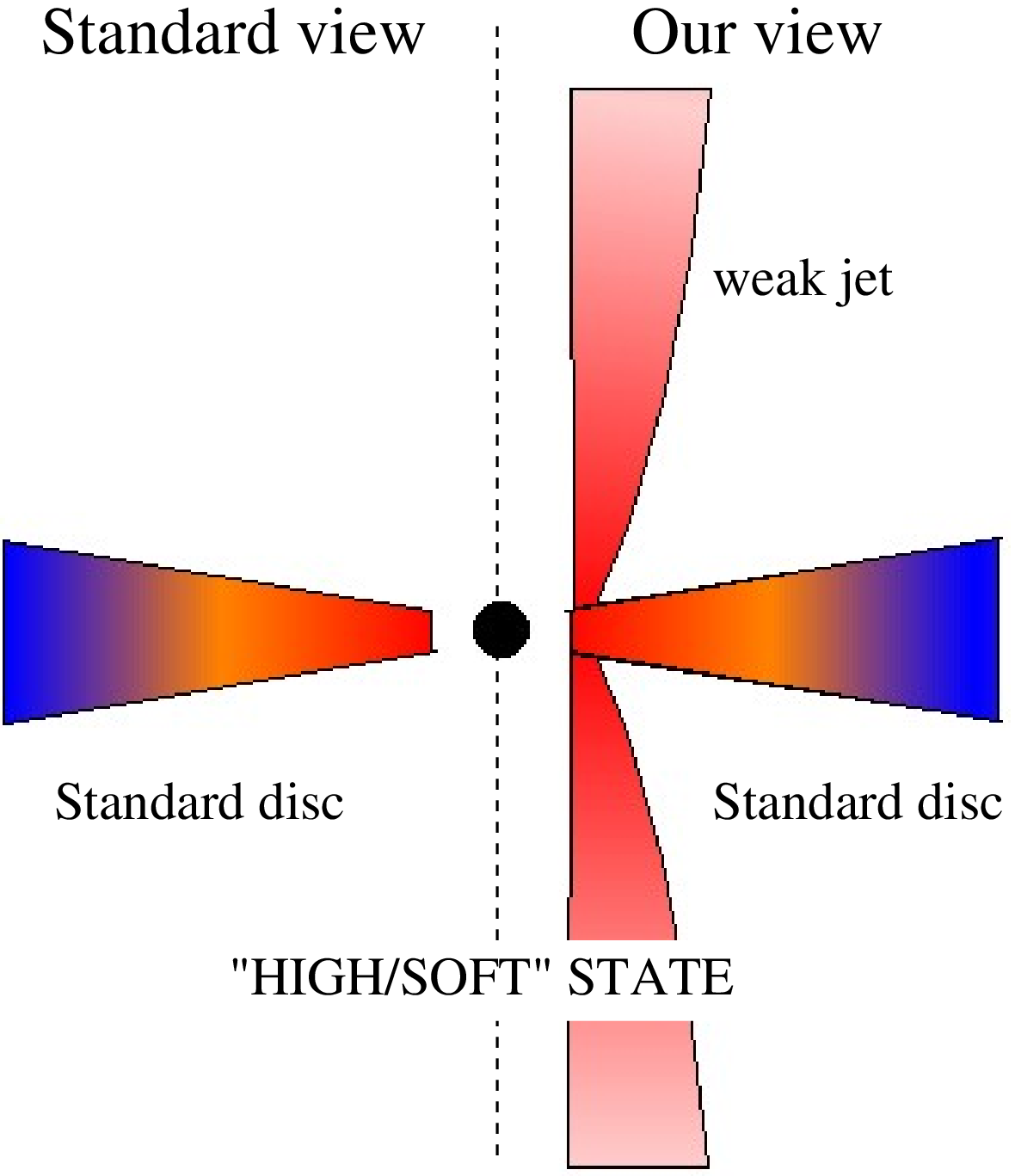}&\hspace*{-4.5cm}\includegraphics[width=0.3\textwidth]{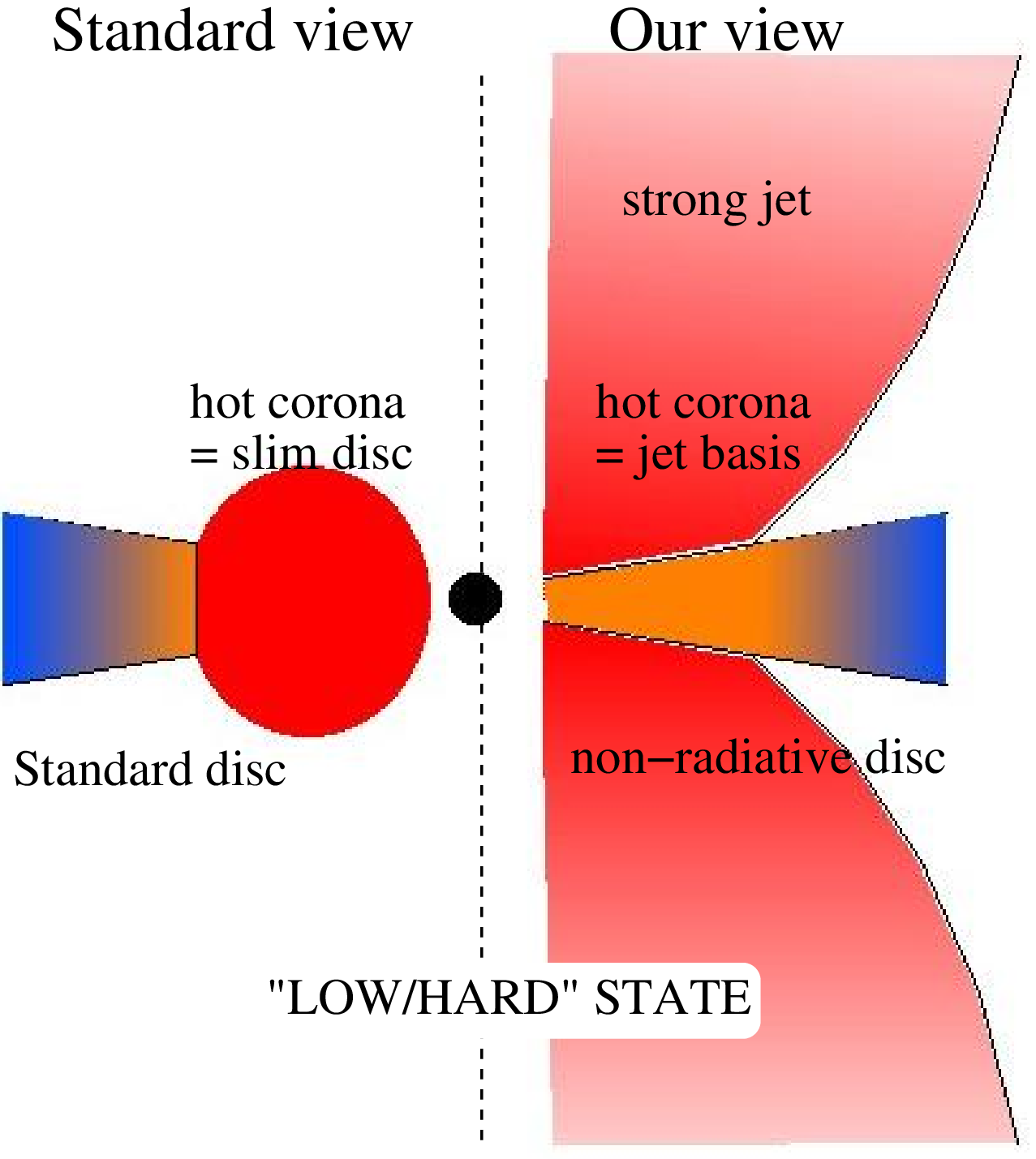}\\
     \hspace*{3cm}\includegraphics[width=0.5\textwidth]{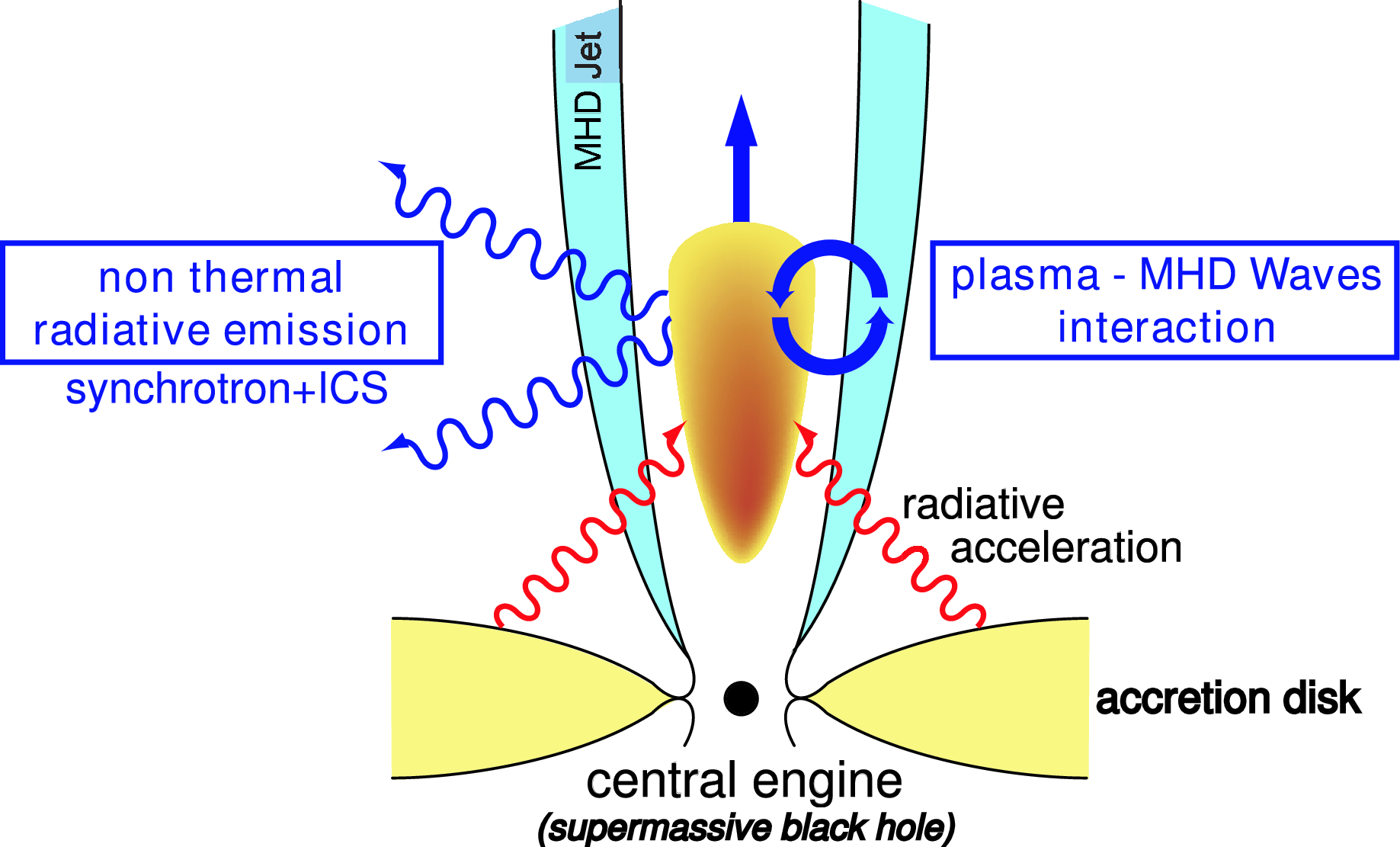}{\begin{minipage}{2cm}
	  \vspace*{-9.cm}\hspace*{-5.cm}
	  \underline{The Two-Flow Model}
      \end{minipage}}& \\
    \end{tabular}
    \caption{Upper left: the high/soft state is characterized by a weak
      jet (weak radio emission) and strong disc emission. Upper right:
      the low/hard state is characterized by a strong jet (strong radio
      emission) and thus weak disc emission. In both state we also
      suppose the existence of a relativistic jet, produced inside the
      MHD one (the two-flow model, lower left. cf. Pelletier 2004 and
      reference therein for more details) and energized by it trough
      plasma-MHD waves interaction. This relativistic jets would produce,
      first the superluminal motion observed in some conditions in
      microquasars, and second it would produces the high energy power
      law tail observed in these states (Ling \& Wheaton 2003). }
    \label{figsketch}
\end{figure}
Jets from microquasars share the property of an extreme collimation with
AGN jets. Such a collimation cannot be achieved by the outer pressure and
requires therefore self-generation. Only large scale magnetic fields that
are carried along with the jet have been proved to provide such a
self-collimation. We assume therefore that a magnetic field is anchored
onto the accretion disk around the compact object giving birth to a MAES.

We aim to explain the two main microquasars spectral states, i.e. the
low/hard and high/soft states, in the MAES framework. Simple cartoons of
our model are shown in Fig \ref{figsketch}. In both states we suppose the
existence of a MAES but the part of the accretion disc that ejects is
much smaller in the high/soft compared to the low/hard state. In this
case, the disc emission is strong and close to the standard one as
observed in this state. Inversely, in the low/hard state the disc
emission is weak due to a strong jet emission (cf. next Sect.). The jet
basis is expected to be relatively hot and could explain the thermal
Comptonized component observed in this state. The strong jet also explain
the strong radio emission associated with this state. As we can see, one
of the main parameter of the model is the radial extension of the
ejection region in the disc. We believe that it is controlled by some
disc instabilities, and for instance part of it may transit from
optically thick to optically thin states.

In this model, we also assume the presence of a relativistic jet produced
inside the MHD one (the two-flow model, Pelletier 2004 and reference
therein) and energized by plasma-MHD waves interaction. This is the
two-flow (outside MHD jet and inside relativistic one) model developed
in our team (Pelletier 2004 and references therein). A zoom of the central
region of the jet is also shown in Fig. \ref{figsketch}. The two-flow
model has been already applied with success to explain the SEDs of AGNs
with jets (e.g. Marcowith et al. \cite{mar98}). In the case of
microquasars, the relativistic jet could explain the superluminal motions
that can be observed in some cases. It could also explain the presence of
the high energy tail generally observed in both states (e.g. Ling \&
Wheaton 2003), and which is not easily explained in the standard view.

We present below the general picture for handing the radiative transfer
problem in MAES.

\section{Magnetized Accretion-Ejection Structures}
\begin{figure}[t]
  \vspace{2mm}
  \begin{center}
    \begin{tabular}{cc}
      \hspace{3mm}\includegraphics[width=0.5\textwidth]{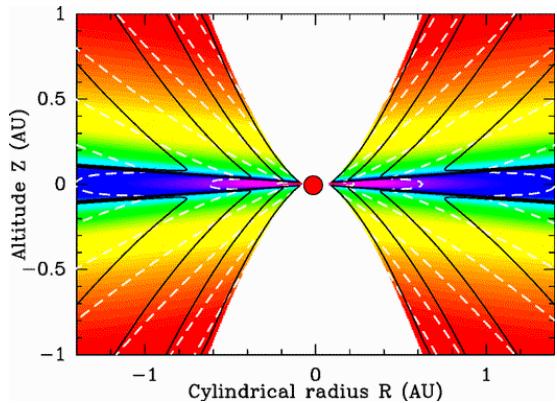}
      \parbox{180mm}{{\vspace{2mm} }}&
      \begin{minipage}{0.45\textwidth}
	\vspace*{-5cm}
	\caption{ Cross-section of an accretion disk driving cold jets
	  (Ferreira 1997). Colors are density contours, streamlines are black
	  solid lines, while white dashed lines show isocontours of total
	  velocity.}
	\label{Fig1}
      \end{minipage}
    \end{tabular}
  \end{center}
\end{figure}
\label{sect:Obs}
In a series of papers (e.g. Ferreira \& Pelletier 1995; Ferreira 1997;
Casse \& Ferreira 2000) it has been shown that, in MAES, the poloidal
magnetic field must be close to equipartition with the plasma thermal
pressure in order to allow for steady ejection. Indeed, in any other
situation, the magnetic field would produce an overwhelming vertical
compression and the plasma pressure would not be strong enough to push
out disk material and load the field lines with matter.

The presence of such an equipartition field has severe consequences on
the disk dynamics since it is very difficult to get rid of the poloidal
magnetic flux.  If this assumption is correct, then every spectral state
of microquasars should provide an observational evidence of its presence.

\begin{figure}[t!]
  \vspace{2mm}
  \begin{center}
      \hspace{3mm}\includegraphics[width=0.7\textwidth]{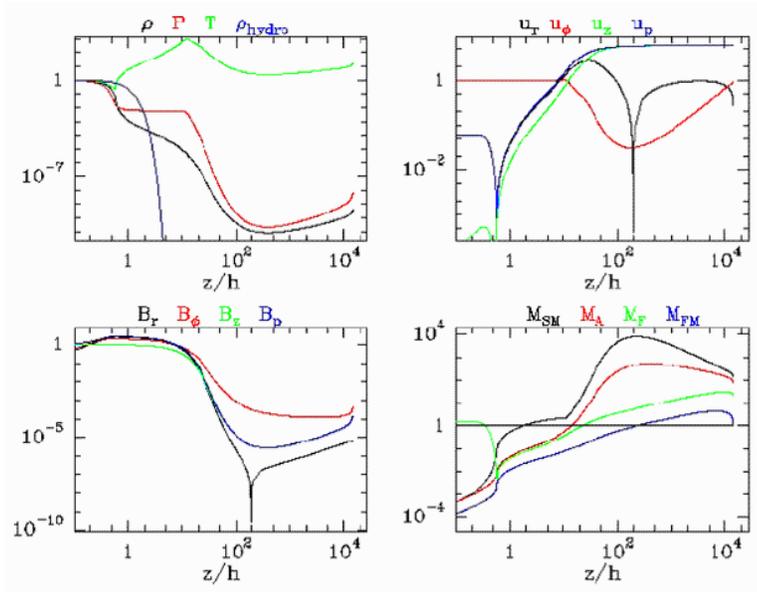}
      \parbox{180mm}{{\vspace{2mm} }}
	\caption{Quantities along a field line for a Self-Similar solution
	  that crosses all three critical points SM, Alfven and FM
	  (Ferreira \& Casse 2004). Density, pressure, temperature and
	  magnetic fields are normalized to their values at the disk
	  mid plane, velocity to the Keplerian speed at the anchoring
	  radius. This solution recollimates at $z\sim 200h$, where $h$
	  is the disc half-height.  Our theory of steady jet formation
	  from Keplerian accretion disks has been confirmed by numerical
	  MHD simulations (Casse \& Keppens 2004).}
	\label{Fig2}
  \end{center}
\end{figure}

Two major characteristics of MAES are:
\begin{itemize}
\item {\bf \underline{Jet torque}:} Magnetically driven jets produce a torque on the
underlying accretion disc such that $\displaystyle
\Lambda=\frac{\mbox{jet torque}}{\mbox{viscous torque}}\sim \frac{r}{h} >
1$. As a consequence, {\bf for the same accretion rate, discs driving jets
are always less dense than standard/ADAF discs}.\vspace*{0.3cm}

\item {\bf \underline{Energy budget}:} The released accretion energy is
shared between the jet power and the disc emission, i.e $\displaystyle
{\mathcal{P}}_{\mbox{acc}}=2{\mathcal{P}}_{\mbox{jet}}+2{\mathcal{P}}_{\mbox{disc}}$
with
$\displaystyle\frac{2{\mathcal{P}}_{\mbox{jet}}}{{\mathcal{P}}_{\mbox{acc}}}=\frac{\Lambda}{1+\Lambda}\simeq
1$. Thus, {\bf jets carry away most of the accretion power and discs
producing jets are weakly dissipative (Ferreira \& Pelletier 1995)}. This
is a key point in our scenario (cf. Sect. 2) since it gives an
alternative to conventional ADAF models, the accretion energy can be
efficiently converted in ejection and not advected inside the
horizon. Thus no (or weak) disc emission does not necessarily mean disc
disappearance but more likely matter ejection! Such result can only be
catched by solving the complete set of MHD equation of the disc-jet
configuration.
\end{itemize}
A vertical cross-section of a MAES is shown in Fig. \ref{Fig1}. The main
parameters in the equations controlling the MAES are $\epsilon = h/r$
where $h$ is the disc half-height, $\mu = B^2/\mu_0 P_{tot}$ and
$m_{acc}= M_{acc}/M_{edd}\propto r^{\xi}$ where the ejection efficiency
$\xi$ can vary between 0.01 up to 0.4. These values are not guessed (like
in the ADIOS model) but were obtained by solving the full set of MHD
equations and scanning the parameter space (Casse \& Ferreira 2000). For
example, as already noted above, a stationary disc+jet solution
corresponds to $\mu\simeq 1$ i.e. equipartition between magnetic field
and plasma pressure.
As a result, we obtain the self-consistent spatial distribution
of the main physical quantities. An example of such distributions along
the line field is plotted in Fig. \ref{Fig2}.

\section{SEDs: preliminary results}
\begin{figure}[b!]
\begin{tabular}{cc}
  \includegraphics[width=0.45\textwidth]{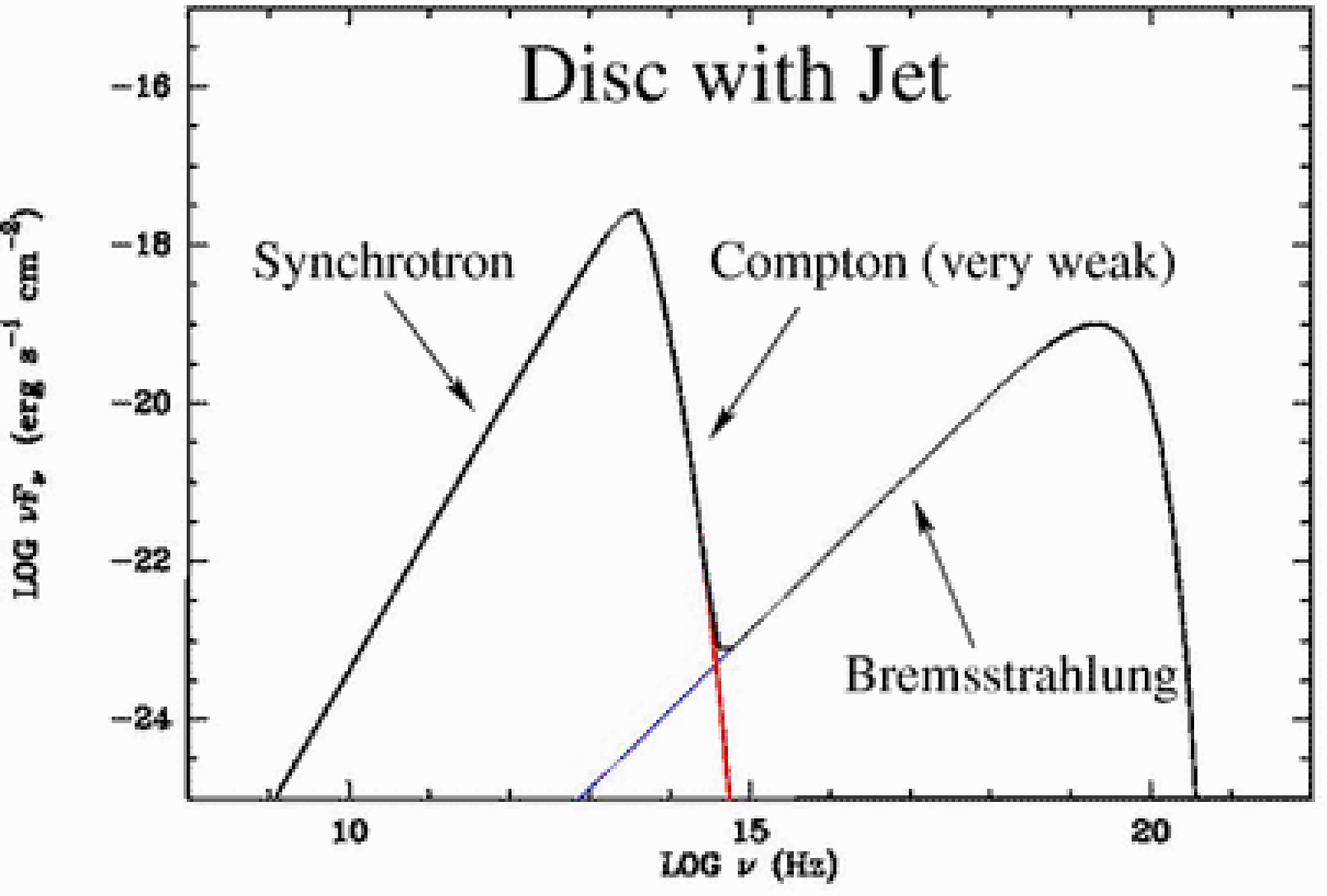}&\includegraphics[width=0.45\textwidth]{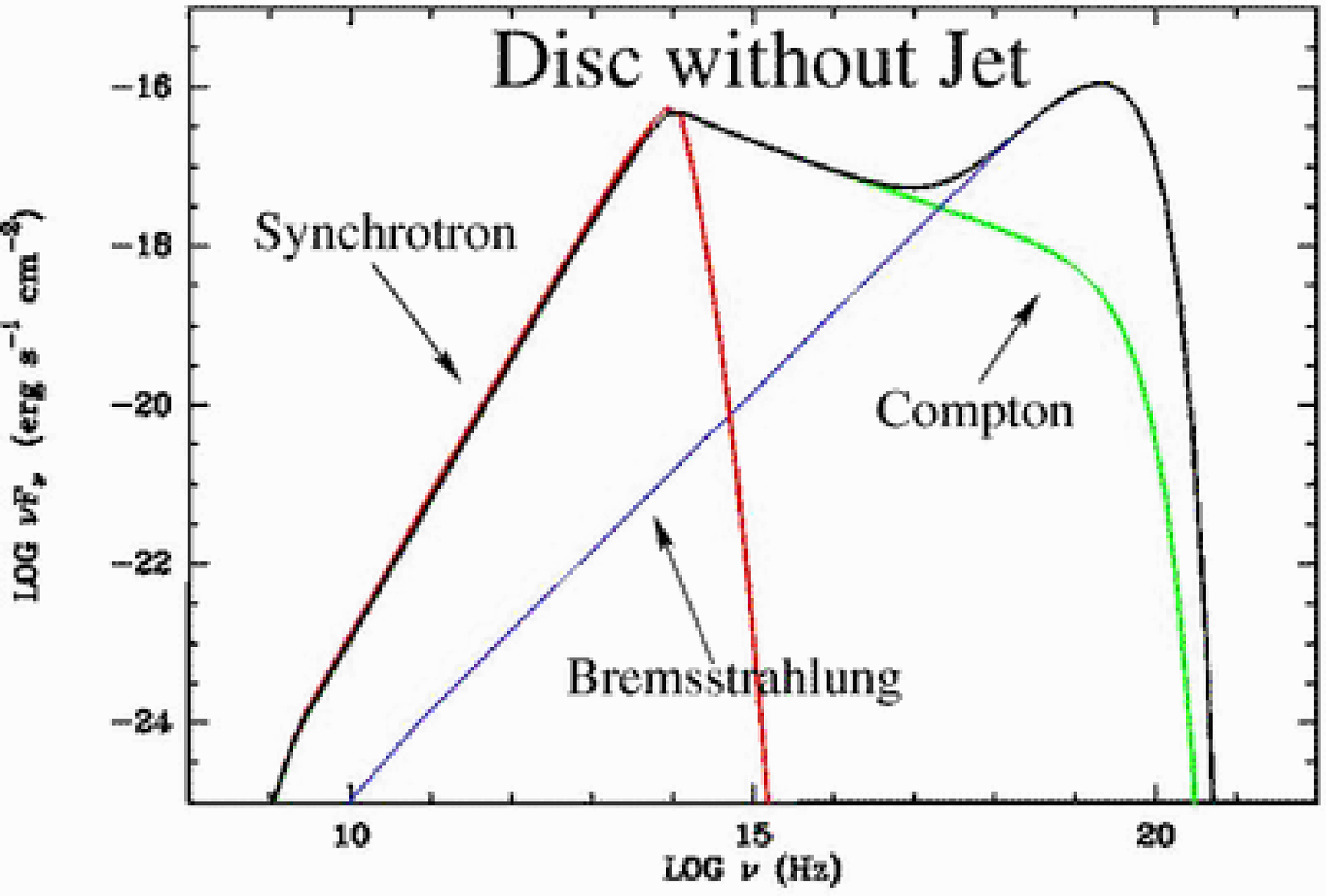}\\
\end{tabular}
  \caption{Two examples of disc SEDs produced by our model with (right)
  and without (left) jet. When there is a jet, the disc density strongly
  decreases (cf. Sect. 3). Consequently, and since in this example we
  assume a fixed electron temperature, the global disc flux weakens with
  a strong decrease of the bremsstrahlung and Compton emissivity. The
  parameters are reported in the text. Red: synchrotron, blue:
  bremsstrahlung, green: Compton. Black: total}
  \label{Fig:plot2}
\vspace{-1cm}
\end{figure}

As already said, we aim to develop a code that computes the total SEDs
emitted by the disc+jet structure. Thus we need to combine the stationary
solutions obtained for MAES (as shown in the previous section) with the
energy equations followed by electrons and protons. This is a work in
progress and at the moment, we compute only the SED produced by the disc
alone, with or without jet emission, for different sets of parameter
values i.e. the energy equation systems is still not solved self
consistently.

We take into account the main radiative processes for the electron i.e
synchrotron, bremsstrahlung (including self-absorption) and Compton
effects.  The disc is divided in a large number of rings and the
different emissivities are computed at each ring, using for the electron
density and magnetic filed the expression consistent with a stationary
MAES. We then sum over the disc surface to obtain the total
emissivity. The figure below shows two examples of SEDs for an electron
temperature $T_e=10^9$ K, $\epsilon = h/r =0.3$, $m_{acc} = 0.01$, $M=10
M_{sun}$ located at 10 kpc. For these parameter values, the disc is
optically thin. We note that, when the energy equations for electrons and
protons will be solved self-consistently, the values of $T_e$ and
$\epsilon$ will be naturally fixed.

The left plot corresponds to the case with jet and the right plot to the
case without jet.  Clearly, the presence of a jet strongly modified the
disc SED!  Indeed, when there is a jet the disc density strongly
decreases (cf. Sect. 3). Consequently, and since we assume a fixed
electron temperature, the global disc flux weakens with a strong decrease
of the bremsstrahlung and Compton emissivity. 

We note that we only want here to exemplify the impact of jet emission on
the disc SED without trying to reproduce the observed microquasar
spectral states. This will be of course the next step of this study.
However, from this simple example, we can extrapolate that if we had
assumed a constant total luminosity instead of a constant temperature,
the case ``disc-with-jet'', would have had a higher electronic
temperature, i.e. a hotter corona, to compensate the smaller
density. This is in agreement with the presence of a hot thermal plasma
commonly assumed in microquasars to explain the X-ray spectra in the
low/hard state.


\section{Conclusion and Perspectives}
We present here only some details of the model we are developing to
explain the spectral changes in microquasars. We believe that the changes
are mainly due to physical changes in the disc itself, and for instance
part of it may be transiting from optically thin to optically thick
states. This would strongly influence the disc-jet structure resulting in
important changes of the global SED.

Work is in progress to develop a complete self-consistent solution of
MAES with realistic radiative transfer in the disc and in the jet.

\end{document}